\documentclass[journal=jpclcd,manuscript=letter]{achemso}

\usepackage[version=3]{mhchem} 
\usepackage{comment}

\author{Mustapha Driouech}
\affiliation{Friedrich-Schiller Universit\"at Jena, Institute for Condensed Matter Theory and Optics, 07743 Jena, Germany}
\alsoaffiliation{Carl von Ossietzky Universit\"at Oldenburg, Institute of Physics, 26129 Oldenburg, Germany}
\author{Michele Guerrini}
\affiliation{Carl von Ossietzky Universit\"at Oldenburg, Institute of Physics, 26129 Oldenburg, Germany}
\author{Caterina Cocchi}
\affiliation{Friedrich-Schiller Universit\"at Jena, Institute for Condensed Matter Theory and Optics, 07743 Jena, Germany}
\alsoaffiliation{Carl von Ossietzky Universit\"at Oldenburg, Institute of Physics, 26129 Oldenburg, Germany}
\alsoaffiliation{Center for Nanoscale Dynamics (CeNaD), 26129 Oldenburg, Germany}
\email{caterina.cocchi@uni-jena.de}

\title{Excited State Absorption Drives Low-Energy Optical Limiting in Oligothiophenes}

\begin{document}


\begin{abstract}
Optical limiting (OL), a crucial mechanism for protecting human eyes and sensitive sensors from intense radiation, relies on understanding the optical nonlinearities acting on the systems. Assessing and disentangling the effects at play is crucial to predict and control the nonlinear optical response in real materials. In this \textit{ab initio} study based on real-time time-dependent density-functional theory, we investigate non-perturbatively the absorption spectra of a set of thiophene oligomers, the building blocks of technologically relevant organic semiconductors, excited by broadband radiation of increasing intensity. Under strong electric fields, the absorption cross section grows significantly below the onset of linear excitations, exhibiting saturation typical of OL. By exciting the oligothiophenes with a train of pulses targeting the first and second excited states of each moiety and analyzing the resulting population dynamics, we reveal excited-state absorption (ESA) in the near-infrared to visible region. Our results indicate ESA as the driving mechanism for OL in oligothiophene molecules, thereby providing important insight to design novel compounds with optimized nonlinear optical characteristics.

\begin{tocentry}
    \centering
    \includegraphics[width=\textwidth]{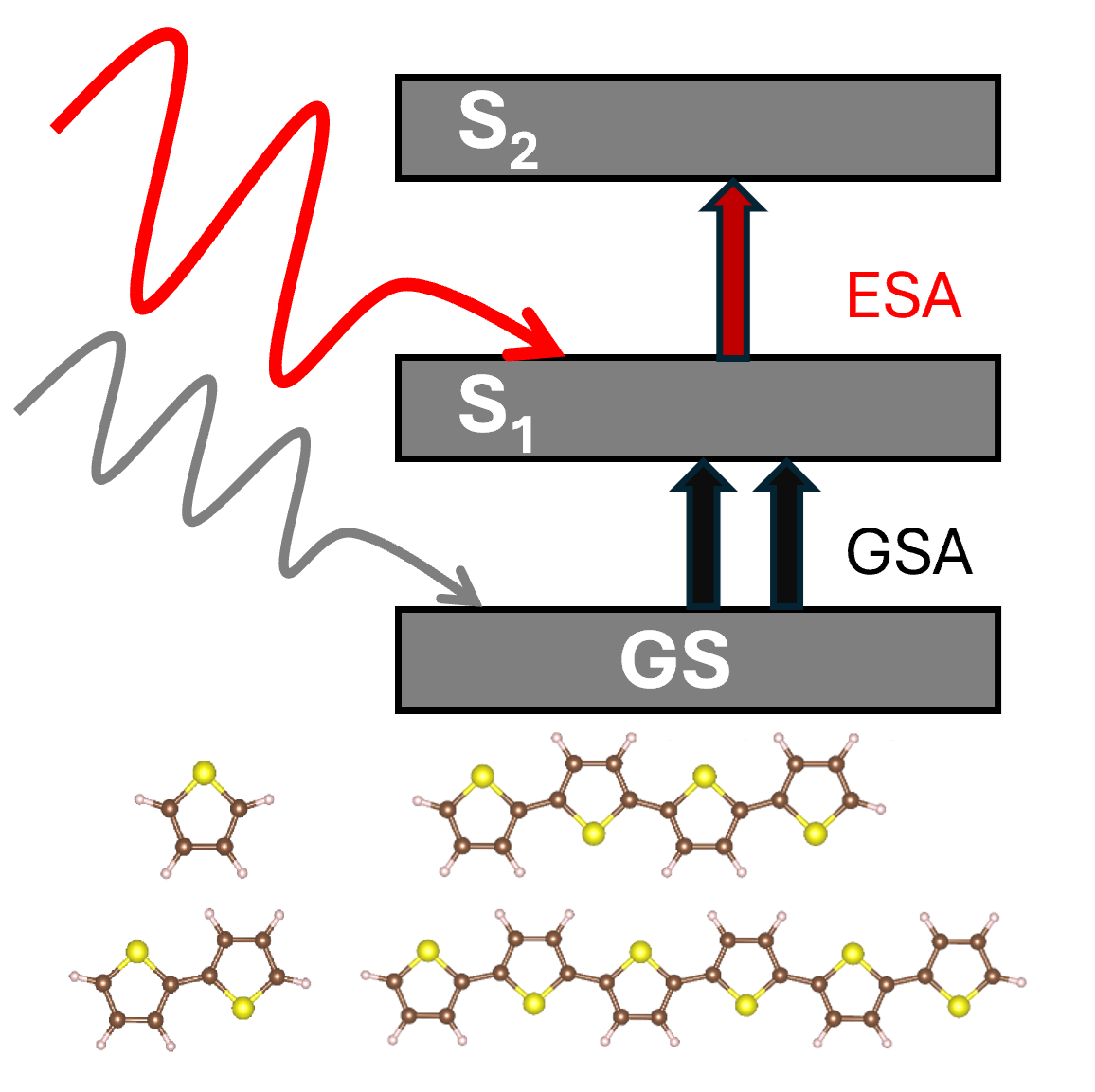}
\end{tocentry}

\end{abstract}


Optical limiting (OL), the ability of a material to attenuate intense light transmission, is of critical importance for laser safety technologies~\cite{reen+22om,saba-abit24acsamn}, including the protection of sensitive optical sensors~\cite{hege+19japs} and human eyes~\cite{grou00om,li+00jnopm,muri+16cap}. Designing efficient OL materials hinges on a fundamental understanding of the underlying optical nonlinearities. While mechanisms like reverse saturable absorption, excited-state absorption (ESA), and two-photon absorption are recognized as primary drivers of OL~\cite{tutt-bogg93pqe}, identifying and disentangling their contributions in specific material classes remains a challenging task. Solving this conundrum is essential for advancing OL-based technologies.

Carbon-conjugated molecules have been intensively studied for OL due to their extended electronic $\pi$-network, large polarizability, and inherent chemical flexibility~\cite{sun-rigg99irpc}. Porphyrins and phthalocyanines have received significant attention, owing to their broad transparency window in the visible region that can be populated by intense radiation~\cite{dela+04cr,calv+04sm}. 
While this interest has significantly promoted the study of OL and the development of applications~\cite{liu+19chemistry}, it has diverted attention from other potentially relevant compounds, such as oligothiophene molecules. Their rich spectrum of linear excitations together with their tunability via chemical functionalization and length modulation~\cite{salz07jctc,bada+10jpcb,krum+21pccp} have established them as building blocks for organic electronic devices~\cite{saka+08oe,xia+08oe,yang+23amr,zaie+23jpca}. Although recent experimental studies on oligothiophene-functionalized graphene~\cite{liu+09carbon,midy+10small} indicate their favorability for nonlinear optics and OL, the potential of oligothiophenes for this type of application remains largely unexplored. A detailed investigation of their response to strong fields is urgently needed to assess their ability as efficient OL compounds. 

\textit{Ab initio} methods are particularly well-suited for studying nonlinear optical properties of molecules. In contrast to empirical models, they do not require any input from experiments, thus representing a reliable and predictive tool to characterize new compounds. Real-time time-dependent density functional theory (RT-TDDFT), a non-perturbative first-principles approach, offers a particularly versatile framework. The ``$\delta$-kick'' method introduced by Yabana and Bertsch to simulate linear absorption spectra~\cite{yaba-bert96prb} and subsequently extended to probe nonlinear excitations driven by intense broadband radiation~\cite{cocc+14prl,fisc+16jpcl,guan+21pccp}, is complemented by efficient schemes for pump-probe~\cite{krum+20jcp} and multidimensional spectroscopy~\cite{krum+24apr} based on the application of pulsed electric fields of tunable shape, frequency, duration, and polarization, evolving with the system during a femtosecond (fs) time window~\cite{degi+13chpch}. This setup enables exploring dynamical charge transfer and non-equilibrium dynamics in organic, inorganic, and hybrid materials~\cite{zhan+17as,lian+18ats,jaco+20apx,ofer+21prl,liu+21prb,uemo+21prb,jaco+22acsanm,urat-naka23jpcl,qi+24jcp}, including the influence of vibronic couplings when combined with Ehrenfest dynamics~\cite{jaco+23jpca,guer+23tca,jaco+24es,xu+24jcp}. %

In this work, we apply the $\delta$-kick and pump-probe schemes of RT-TDDFT to investigate OL in four thiophene oligomers composed of 1, 2, 4, and 6 rings. By exciting the molecules with broadband radiation of increasing intensity, we find enhanced nonlinear absorption in the near-infrared to visible region below the onset of the linear spectrum. The saturation of this band upon increasing field intensity confirms its relation to OL. By impinging the oligothiophenes with a train of fs pulses and analyzing the resulting population dynamics, we rationalize their nonlinear optical behavior in terms of ESA. Our results have two important implications: they disclose the potential of oligothiophenes for OL in the near-infrared to visible region, depending on their length, and confirm the ability of RT-TDDFT to shed light on optical nonlinearities of conjugated molecules in an insightful and yet computationally efficient way.

\begin{figure}[h!]
  \includegraphics[width=\linewidth]{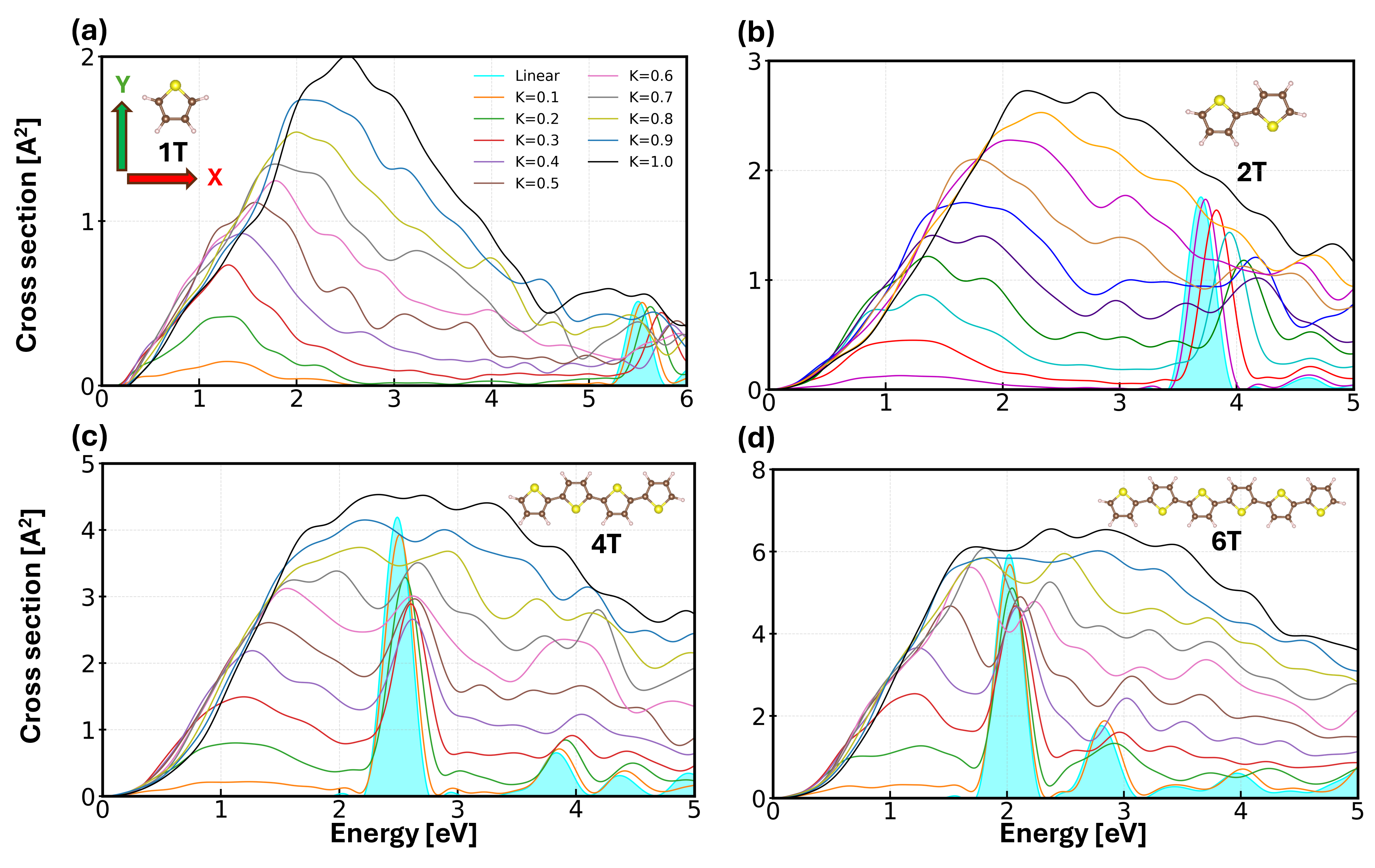}
  \centering
  \caption{Absorption cross section computed for a) 1T, b) 2T, c) 4T, and d) 6T excited by instantaneous, broadband excitations of increasing intensities. Linear absorption spectra are obtained with a $\delta$-kick of 0.001~\AA{}$^{-1}$ (cyan filled areas), while nonlinear spectra are excited by kicks $\kappa$ in \AA{}$^{-1}$ specified in the legend of panel a). Insets: ball-and-stick representations of the investigated oligothiohene molecules with C atoms depicted in brown, S atoms in yellow, and H atoms in pink. The Cartesian coordinate system referred to all molecules is visualized in panel a).}
  \label{fgr:kick_fig1}
\end{figure}

We start our analysis by computing the absorption spectra of the four considered oligothiophenes applying $\delta$-kicks of increasing intensity, ranging from a weak perturbation delivering the linear spectrum~\cite{yaba-bert96prb} to strong fields triggering pronounced nonlinear response~\cite{cocc+14prl}. The linear spectrum of the single thiophene ring (1T), obtained with a $\delta$-kick of magnitude 0.001~\AA{}$^{-1}$, exhibits a peak around 5.5~eV (Figure~\ref{fgr:kick_fig1}a), which, due to intrinsic broadening (details in the Computational section) encompasses the two lowest-energy excitations polarized in the $y$- and $x$-direction, respectively~\cite{cocc-drax15prb} (see Table~S1 for the perturbative analysis of the linear excitations). Our findings are in very good agreement with quantum chemistry predictions~\cite{salz+08pccp,koel+16pccp} and experimental results~\cite{holl+14pccp,habe+03pccp}. Increasing the perturbation strength by two orders of magnitude ($\delta$-kick $\kappa=0.1$~\AA{}$^{-1}$) preserves the first absorption peak around 5.5~eV, but also leads to a non-zero cross section at low energies, between 0.5 and 2.5~eV. Larger values of the $\delta$-kick induce a blue shift in the peak associated with the first linear excitation around 5.5~eV, a consequence of memory effects being neglected in the adopted adiabatic local density approximation~\cite{wije-ulri08prl,fuch+11prb}. Importantly, the low-energy absorption band grows in intensity and further extends in energy with the magnitude of the kick, merging with the peak at 5.5~eV for $\kappa \geq 0.8$~\AA{}$^{-1}$ (Figure~\ref{fgr:kick_fig1}a). 

A similar behavior is exhibited by bithiophene (2T) under analogous excitation conditions. A weak $\delta$-kick $\kappa = 0.001$~\AA{}$^{-1}$ leads to the linear absorption spectrum of this molecule, characterized by a sharp resonance around 3.8~eV (Figure~\ref{fgr:kick_fig1}b), in overall agreement with experiments~\cite{habe+03pccp,lap+97jpca} and quantum-chemistry calculations~\cite{andr-wite11tca}. Increasing the field intensity induces again a slight blue shift of the main absorption maximum due to the lack of memory effects in our calculations~\cite{wije-ulri08prl,fuch+11prb}, and the appearance of a broad absorption band centered at 1~eV (Figure~\ref{fgr:kick_fig1}b). An even stronger perturbation shifts the maximum to 4~eV and further extends the energy range of the low-energy  band across the entire visible range. 

The linear absorption spectra of quaterthiophene (4T) and sexithiophene (6T) are characterized by strong resonances in the visible region, centered at approximately 2.5~eV (Figure~\ref{fgr:kick_fig1}c) and 2~eV (Figure~\ref{fgr:kick_fig1}d), respectively. These findings are in line with previous \textit{ab initio} calculations~\cite{leng+15jcp,sun+15pps} and experiments~\cite{birn+92jcp,yang+98jpcb,rent+99pccp}. In the spectrum of 6T, a second weaker maximum appears at 2.8~eV. For 4T, this excitation has a lower oscillator strength and a higher energy, appearing at approximately 4 eV. In analogy with the shorter oligomers, perturbing these molecules with $\delta$-kicks of increasing intensity promotes absorption in the low-energy spectral region, corresponding to infrared frequencies. The first absorption peak in the linear regime remains well-defined up to $\kappa=0.4$~\AA{}$^{-1}$, despite losing oscillator strength and being slightly blue-shifted (Figure~\ref{fgr:kick_fig1}c,d). In the spectrum of 4T, larger $\delta$-kick intensities, up to $\kappa=0.7$~\AA{}$^{-1}$ enhance the spectral strength of this resonance while broadening it, while for $\kappa>0.7$~\AA{}$^{-1}$, a continuous absorption band rising at approximately 0.5~eV up to 4~eV is formed (Figure~\ref{fgr:kick_fig1}c). For 6T, the situation is more faceted. Kick strengths between 0.6~\AA{}$^{-1}$ and 0.8~\AA{}$^{-1}$ generate a two-peak structure in the absorption spectrum (Figure~\ref{fgr:kick_fig1}d), while, similar to 4T, stronger intensities give rise to featureless absorption from infrared to near-UV frequencies. The second absorption peak in the linear regime is more sensitive to the kick strength in the spectra of both molecules, where it is no longer distinguishable from $\kappa=0.5$~\AA{}$^{-1}$ in 4T (Figure~\ref{fgr:kick_fig1}c) and $\kappa=0.6$~\AA{}$^{-1}$ in 6T (Figure~\ref{fgr:kick_fig1}d).

The large low-energy absorption cross-section in the nonlinear spectra of the considered thiophene oligomers suggests the emergence of OL, further confirmed by its saturation upon integration in the relevant energy window (Figure~S1). In contrast to phthalocyanine, where the spectral window populated by intense, broadband radiation is identified between the two main absorption bands~\cite{cocc+14prl} hosting dark transitions, enhanced nonlinear absorption in the spectra of oligothiophenes appears below the lowest-energy excitation in the linear regime. This finding suggests that ESA drives the response of these molecules to intense broadband radiation. To test this hypothesis, we perform an additional set of RT-TDDFT simulations, exciting the molecules with time-dependent pulses. By targeting the lowest-energy excitation (S$_0 \rightarrow$ S$_1$, see Tables~S1-S4 in the Supporting Information and the Computational Section below), we drive the molecules out of the linear regime. Next, we probe the population dynamics by impinging the moieties with a second pulse, delayed by 15~fs with respect to the first one and with carrier frequency in resonance with the low-absorption band emerging in the nonlinear regime.

\begin{figure}[h!]
  \includegraphics[width=\linewidth]{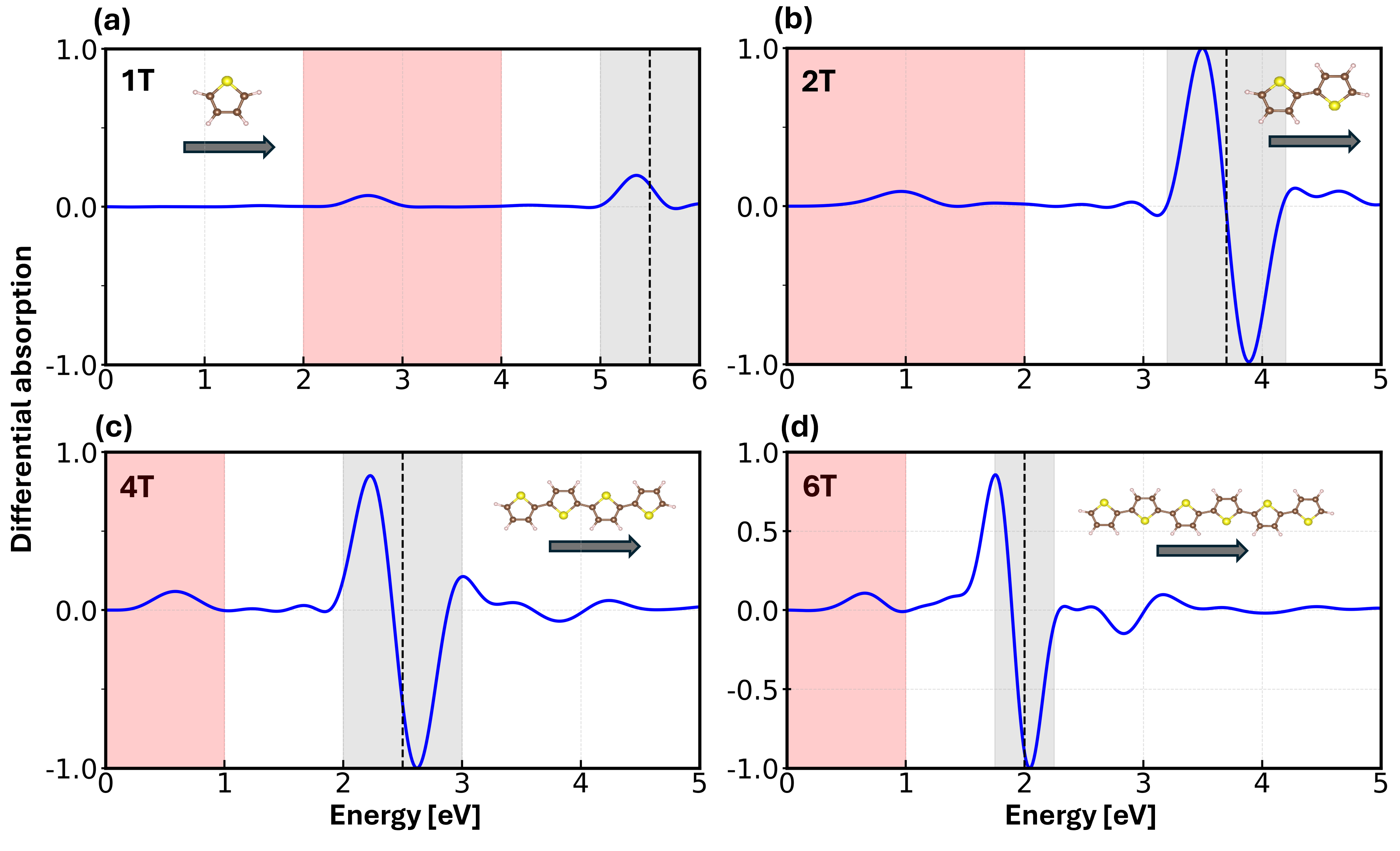}
  \centering
  \caption{Differential absorption spectra of a) 1T, b) 2T, c) 4T, and d) 6T excited by a Gaussian pulse with peak intensity $I=50$~GW/cm$^2$ and photon energy of a) 5.5~eV, b) 3.7~eV, c) 2.5~eV, and d) 2.0~eV, in resonance with the first excitation in the linear spectra of the oligomers (dashed bars). The gray area indicates the bandwidth of the applied pulse, while the red shaded region highlights the low-energy nonlinear absorption band subsequently probed by the second pulse to access the population dynamics.}
  \label{fgr:TAS}
\end{figure}

The differential absorption spectra computed for each molecule after the application of the first pulse targeting S$_0 \rightarrow$ S$_1$ are dominated by the excitation in resonance with the carrier frequency of the applied pulse (dashed vertical line in Figure~\ref{fgr:TAS}). In addition, a weak but distinct absorption peak appears in the low-energy region of each spectrum, in the same window where strong $\delta$-kicks give rise to non-zero absorption cross section (Figure~\ref{fgr:kick_fig1}). For 1T, this maximum appears in the visible region at 2.7~eV (Figure~\ref{fgr:TAS}a), while for the longer oligomers, it is found at infrared frequencies, around 1~eV in 2T (Figure~\ref{fgr:TAS}b), and close to 0.5~eV for both 4T (Figure~\ref{fgr:TAS}c) and 6T (Figure~\ref{fgr:TAS}d). The differential absorption spectra of the two longest molecules, 4T and 6T, are characterized by many more features compared to those of the shorter moieties. This is due to the higher density of excited states, which are involved in the nonlinear excitation. Nonetheless, all nonlinear spectra exhibit the same key characteristic, namely an absorption band in the near-infrared region, which is compatible with ESA. 

\begin{figure}[h!]
  \includegraphics[width=\linewidth]{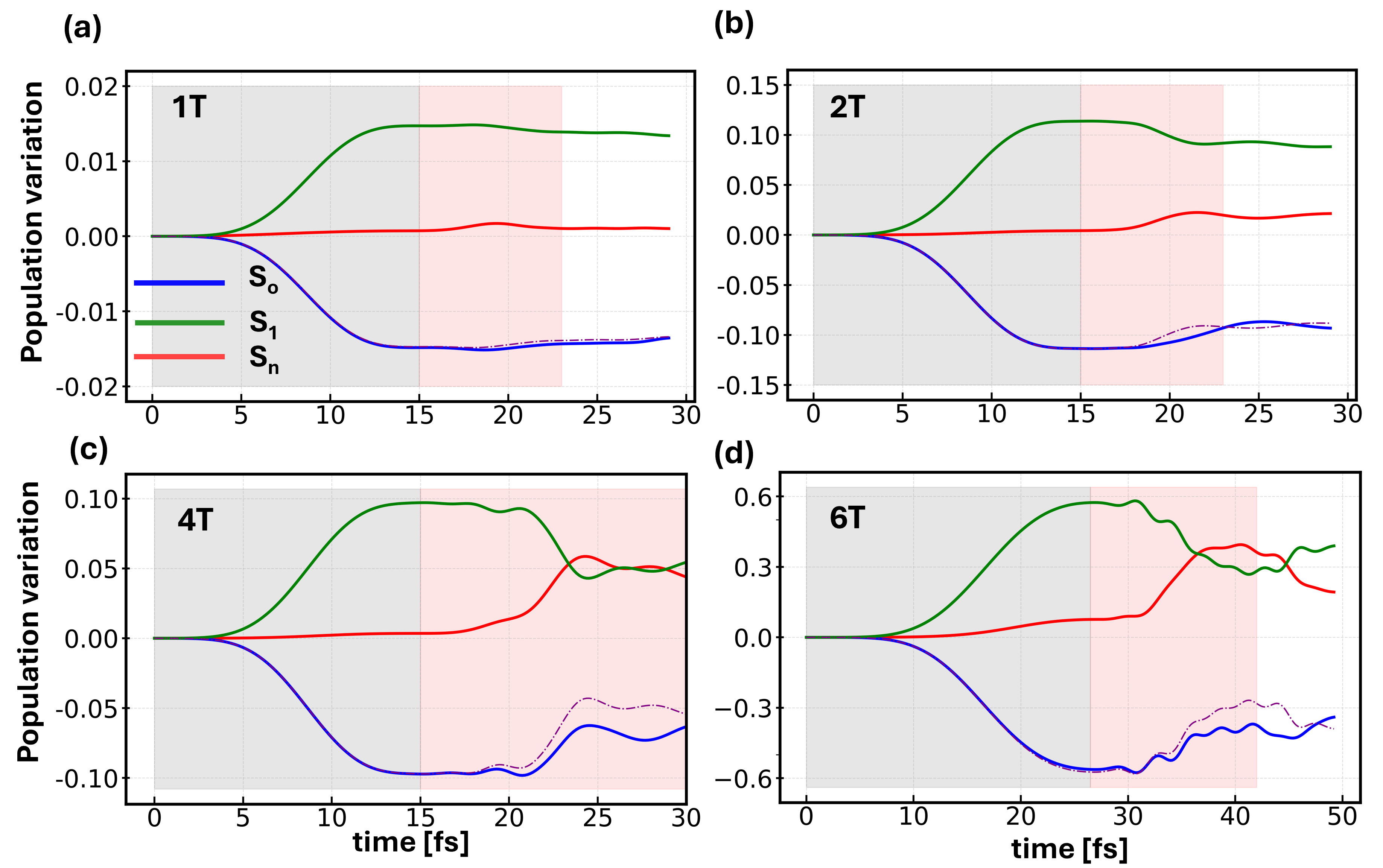}
  \centering
  \caption{Variation in time of the population of the ground state (S$_0$) as well as of the first (S$_1$) and higher (single) excited states (S$_n$) of a) 1T, b) 2T, c) 4T, and d) 6T, driven by a train of two Gaussian pulses in resonance with the S$_0 \rightarrow$ S$_1$ excitation (gray area) and with the absorption maximum below the linear onset (red area). The dotted-dashed purple line indicates the inverse of the population variation of S$_1$.  
  }
  \label{fgr:Popul}
\end{figure}

To confirm this hypothesis, we monitor the electronic population dynamics by estimating the number of excited electrons through the contributions from time-dependent single-particle states $\phi_j(t)$, see Computational Section below, projected onto their ground-state counterparts at $t=0$~\cite{krum+20jcp}:
\begin{equation}
       N_{\text{ex}}(t) = 2 \sum_{m }^\text{unocc} \sum_{j }^\text{occ} \left| \langle \phi_m(0) \mid \phi_j(t) \rangle \right|^2,
    \label{eq:N_ex}
\end{equation}
where the pre-factor 2 accounts for spin degeneracy. We perform this analysis in two steps. Upon the application of the first pulse (gray area in Figure~\ref{fgr:Popul}), which is set in resonance with the S$_0 \rightarrow$~S$_1$ excitation in each molecule (gray area in Figure~\ref{fgr:TAS}), we evaluate the amount of charge depleted from the ground state and promoted to the first excited state. In the single-particle framework provided by RT-TDDFT, we compute these contributions in terms of the occupation of the highest-occupied molecular orbital (HOMO) and the lowest-unoccupied molecular orbital (LUMO) involved in this transition, see Tables~S1-S4. Next, we apply a second pulse targeting the low-energy absorption maximum (red area in Figure~\ref{fgr:Popul}) and estimate the number of electrons promoted from S$_1$ to higher excited states S$_n$, including contributions from LUMO+1 up to LUMO+10, see Figure~S2. 

The results visualized in Figure~\ref{fgr:Popul} validate our hypothesis regarding the role of ESA in driving low-energy nonlinear absorption in the oligothiophene molecules excited by strong broadband radiation. In 1T, the application of the first pulse with a carrier frequency of 5.5~eV triggers the occupation of S$_1$ at the expense of S$_0$ (Figure~\ref{fgr:Popul}a, gray area). After 15~fs, when the second pulse with a carrier frequency of 3~eV is activated (Figure~\ref{fgr:Popul}a, red area), we notice a slight reduction in S$_1$ due to the population of higher excited states (red curve). This is a signature of ESA. Concomitantly, the ground-state population rises, but without perfectly mirroring the S$_1$ population (dashed-dotted purple curve in Figure~\ref{fgr:Popul}a), as during the action of the first pulse. We attribute this behavior to ground-state bleaching during the population of excited states beyond S$_1$.

In 2T, the mechanisms described above for 1T are amplified both quantitatively (compare the $y$-axis scale in Figure~\ref{fgr:Popul}a and Figure~\ref{fgr:Popul}b) and qualitatively. The first pulse with a carrier frequency of 3.7~eV shifts more than 0.1~$e$ from the ground state to S$_1$. After 20~fs, when the second pulse has reached its peak, the population of the first excitation decreases at the advantage of higher states (Figure~\ref{fgr:Popul}b, red area). Again, signatures of a slight ground-state bleaching are visible through the mismatch between the blue solid curve and the dashed-dotted purple line in Figure~\ref{fgr:Popul}b. 

The situation is more faceted for 4T (Figure~\ref{fgr:Popul}c). Here, the activation of the second pulse leads to a sizeable depletion of S$_1$, which becomes less occupied than the other excited states after the second pulse has reached its peak. The S$_n$ population receives non-negligible contributions also from the ground state, as indicated by the blue curve departing from the purple one, representing the inverse population of S$_1$. In 6T, occupation variations become even larger in magnitude (Figure~\ref{fgr:Popul}d) with S$_1$ losing more than 50\% of the population gained from the first pulse during irradiation with the second. Higher excited states take up more than 30\% of the total electronic population upon the application of the second pulse, with ground-state bleaching contributing to the process. It is worth noting that for 6T, excited states beyond S$_1$ are populated already by the first pulse (Figure~\ref{fgr:Popul}d, gray area). The more elaborate composition of the first excited state in this long oligomer (see Table~S4) likely plays a role here. However, we cannot exclude additional nonlinear effects emerging in the laser-driven dynamics, which will be investigated in follow-up work. 

In summary, our RT-TDDFT simulations unambiguously attribute to ESA the absorption below the linear onset induced by intense, broadband radiation in the spectra of 1T, 2T, 4T, and 6T. This nonlinearity is associated with OL and discloses the potential of these molecules to be employed as active components for corresponding applications working in the near-infrared to visible region. We revealed this mechanism by exciting the systems with a train of two pulses tuned in resonance with the first linear excitation in each molecule (S$_0 \rightarrow$ S$_1$) and with transitions from the S$_1$ to higher excited states. The resulting electronic population dynamics support this interpretation, revealing ground-state depletion and the subsequent occupation of the first excited state under the action of the first pulse, as well as the population of higher excited states when the second pulse is turned on. While the length of the molecules and the consequent electronic-structure variations induce expected quantitative changes, the same qualitative behavior persists in the entire series, confirming ESA as the key mechanism driving OL in these compounds.

In conclusion, this study reveals the significant potential of oligothiophene molecules as OL compounds, featuring an active region in the near-infrared to visible band that effectively complements the window covered by more established materials like phthalocyanines and porphyrins. Our detailed analysis based on RT-TDDFT identifies ESA as the main driver of OL in these molecules. This work not only highlights the capability of this parameter-free, non-perturbative \textit{ab initio} method to efficiently simulate and unveil the fundamental origins of optical nonlinearities in conjugated molecules but also sets the stage for the rational design of optimized OL compounds with in-depth insight into the underlying physical mechanisms.

\section{Computational Methods}

\subsection{Theoretical Background}

The calculations presented in this work are based on RT-TDDFT, based on the time propagation of the time-dependent Kohn-Sham (TDKS), 
\begin{equation}
i \hbar \frac{\partial}{\partial t} \psi_i(\mathbf{r}, t) = \left( -\frac{\hbar^2}{2m} \nabla^2 + V_{\text{eff}}(\mathbf{r}, t) \right) \psi_i(\mathbf{r}, t),
\label{eq:TD-KS}
\end{equation}
where $\psi_i(\mathbf{r}, t)$ are the TDKS states and $V_{\text{eff}}(\mathbf{r}, t)$ is the time-dependent effective potential, including the contributions from the external potential, accounting for electron-nuclear interactions, the Hartree potential, and the exchange-correlation potential. The time-dependent electron density, computed from the solution of the TDKS as $\rho(\mathbf{r}, t) = \sum_i^{occ} |\psi_i(\mathbf{r}, t)|^2$, enters the expression of the time-dependent transition dipole moment, which in the $x$-direction reads:
\begin{equation}
\langle x(t) \rangle = \int x \, \rho(\mathbf{r}, t) \, d\mathbf{r}.
\label{eq:tdipole}
\end{equation}
The Fourier transform of Eq.~\eqref{eq:tdipole},
\begin{equation}
\langle\tilde{x}(\omega) \rangle= \int \langle x(t) \rangle e^{i \omega t} \ dt,
\end{equation}
is proportional to the polarizability $\alpha$, whose imaginary part enters the expression of the absorption cross section~\cite{yaba-bert96prb}.

To compute the nonlinear response, we employed two computational approaches.
In the so-called $\delta$-kick scheme, introduced by Yabana and Bertsch~\cite{yaba-bert96prb} and later applied by Cocchi \textit{et al.} to simulate OL~\cite{cocc+14prl}, the system is excited by an instantaneous broadband electric field, causing the electronic wave functions to acquire a phase factor expressed as $ \psi_i(\mathbf{r}, t) \rightarrow \psi_i(\mathbf{r}, t) e^{i \kappa \cdot x} $ in the length gauge. 
The associated electric field is 
\begin{equation}
\mathbf{E}(t) = \mathbf{E}_0 \delta(t),
\end{equation}
with the amplitude $\mathbf{E}_0 = \dfrac{\hbar k}{e}$ depending linearly on the kick strength $\kappa$. 
The molecules are also excited with a time-dependent Gaussian-enveloped electric field of the form
\begin{equation}
\mathbf{E}(t) =  \mathbf{E}_0 \exp\left[-\frac{(t - t_0)^2}{2\tau^2} \right] \cos(\omega_0 t),
\end{equation}
where $\mathbf{E}_0$ is the peak amplitude, $\omega_0$ the carrier frequency, $t_0$ the pulse center time, and $\tau$ the standard deviation related to the full-width half maximum of the Gaussian pulse indicated as the bandwidth of the laser in Figure~\ref{fgr:TAS}. 

\subsection{Computational Details}
All calculations reported in this work were performed with the code \texttt{octopus}~\cite{tanc+20jcp}, implementing RT-TDDFT on real-space numerical grids. We adopted Troullier-Martins pseudopotentials~\cite{trou-mart91pr} and the adiabatic local density approximation as implemented in the Perdew-Zunger functional~\cite{perd-zung81pr}.
The $\delta$-kick simulations were carried out in a real-space box of size 15~\AA{} and grid spacing of 0.18~\AA{}. The TDKS equations were propagated using the enforced time-reversal symmetry propagator and Lanczos algorithm~\cite{lanc50jrnbs} with a time step of 10$^{-3}$~fs = 1~as for a total duration of 15~fs, giving rise to an intrinsic broadening of 44~meV in the linear absorption spectra. The adopted $\delta$-kicks were polarized in all three Cartesian directions to capture the full spectral response of the molecules. The electronic contributions to each linear excitation were resolved using the Casida method~\cite{casi+98jcp} (Tables~S1-S4).

The runs with the Gaussian-shaped electric fields were carried out in a spherical box with a radius of 5~\AA{} and grid spacing of 0.25~\AA{}. The time-step is set to 2.9~as. 
To calculate the differential absorption (Figure~\ref{fgr:TAS}), we excited 1T, 2T, and 4T with a pulse of peak intensity $I=50$~GW/cm$^2$, while for 6T we took $I=10$~GW/cm$^2$. The carrier frequencies were set in resonance with the S$_0 \rightarrow$ S$_1$ transition of each molecule, namely 5.5~eV for 1T, 3.7~eV for 2T, 2.5~eV for 4T, and 2~eV for 6T. To amplify variations in the population dynamics (Figure~\ref{fgr:Popul}), we applied the second pulse with a peak intensity of 2~TW/cm$^2$ for 1T and 2T, 1~TW/cm$^2$ for 4T, and 500~GW/cm$^2$ for 6T. The carrier frequency was set to 3~eV, 1~eV, 0.5~eV, and 0.5~eV for 1T, 2T, 4T, and 6T, respectively, targeting the absorption peak emerging in the differential absorption displayed in Figure~\ref{fgr:TAS}. 

\begin{acknowledgement}
C.C. acknowledges fruitful discussions with Carlo A. Rozzi in the preliminary stage of this project. This work was funded by the German Research Foundation, Project number 524452181, by the State of Lower Saxony (Professorinnen f\"ur Niedersachsen, DyNano, and ElLiKo), and by the Federal Ministry of Education and Research (Professorinnenprogramm III).  Computational resources were provided by the North-German Supercomputing Alliance (NHR), project nip00074.
\end{acknowledgement}

\section{Data Availability Statement}
The data generated in this study, including the corresponding input files, are available free of charge on Zenodo, DOI: https://doi.org/10.5281/zenodo.16037460 .

\begin{suppinfo}
In the Supporting Information, we report the integrated cross sections of the thiophene oligomers, the Casida analysis of their linear excitations, and the molecular orbital contributions to the population analysis.
\end{suppinfo}


\providecommand{\latin}[1]{#1}
\makeatletter
\providecommand{\doi}
  {\begingroup\let\do\@makeother\dospecials
  \catcode`\{=1 \catcode`\}=2 \doi@aux}
\providecommand{\doi@aux}[1]{\endgroup\texttt{#1}}
\makeatother
\providecommand*\mcitethebibliography{\thebibliography}
\csname @ifundefined\endcsname{endmcitethebibliography}
  {\let\endmcitethebibliography\endthebibliography}{}

\end{document}